\documentclass[twocolum,12pt]{iopart}

\usepackage{amsfonts}
\usepackage{bbold}
\usepackage{mathrsfs}
\usepackage{color}
\usepackage{epstopdf}

\usepackage{graphicx}
\usepackage{bm}
\usepackage{epsfig}

\usepackage{amssymb, dsfont}

\newcommand{\nn}{\nonumber }

\newcommand{\rr}{{\mathbf r}}

\newcommand{\BEQ}{\begin{equation}}
\newcommand{\EEQ}{\end{equation}}
\newcommand{\BEA}{\begin{eqnarray}}
\newcommand{\EEA}{\end{eqnarray}}

\begin{document}

\title {Run-and-tumble particles in speckle fields}
\author{M. Paoluzzi, R. Di Leonardo, and L. Angelani}
\address{CNR-IPCF, UOS Roma, Dip. Fisica, Universit\`a {\em Sapienza}, P. le A. Moro 2, I-00185,
  Rome, Italy}
\ead{matteo.paoluzzi@ipcf.cnr.it}



\begin{abstract}
The random energy landscapes developed by speckle fields 
can be used to confine and manipulate a large number of micro-particles with a single laser beam. 
By means of molecular dynamics simulations, we investigate the static and dynamic properties of an
active suspension of swimming bacteria embedded into speckle patterns. 
Looking at the correlation of the density fluctuations and the equilibrium density profiles, 
we observe a crossover phenomenon when the forces exerted by the speckles are equal to the bacteria's propulsion.
\end{abstract}

\maketitle

\section{Introduction}
Starting from the seminal paper of Ashkin \cite{Ashkin70},
optical trapping has developed into a powerful technique, 
widely used in many scientific areas, to manipulate atoms \cite{Ashkin97},
Bose-Einstein condensate \cite{Stamper-Kurn98}, viruses and bacteria \cite{Ashkin87}.
By means of holographic optical tweezers it is possible to trap array of particles
or molecules in three dimensions \cite{Grier03,Padgett11}.
More recently it has been demonstrated that a static speckle pattern, generated by
the interference of random coherent wavefronts, can trap and manipulate a large number 
of particles in three dimensions \cite{Shvedov10}.
Brownian motion in
random energy landscapes provides useful models
to study theoretically and experimentally different phenomena
like anomalous transport in inhomogeneous media \cite{Sokolov12,Bouchaud90,Dean07,Zwanzig88,Havlin87,Isichenko92}, the relaxation
properties of disordered and glassy materials \cite{Jack09,Heur08,Bernasconi79,Bouchaud90}, anomalous diffusion in
living matter \cite{Barkai12} and in disordered media \cite{Haus82,Novikov11}. Colloids in one 
\cite{Hanes12a,Hanes12b} and two dimensional \cite{Evers13} random energy
landscapes have been recently investigated in experiments and by means of numerical simulations \cite{Volpe14}.
While many efforts have been
devoted to study passive particles in random potentials, the behavior of
active objects has been only recently explored \cite{Chepizhko13,Reichhardt14}.

We investigate the dynamics of active particles in the energy landscape provided by speckle patterns.
The microscopic dynamics that we will address, namely run-and-tumble \cite{Schnitzer93}, mimics the motion of swimming bacteria as {\itshape E. coli} \cite{Berg04,Berg72,Koumakis13,DiLeonardo10}.
Run-and-tumble is a simple but powerful 
model that captures many properties of motile bacteria \cite{Schnitzer93,Cates12,Reichhardt13,Reichhardt08}. 
From the theoretical point of view,  
in the non-interacting 
limit (``ideal gas'' of active particles with no steric interactions), density fluctuations can be computed analytically
in one, two and three dimensions \cite{Angelani13,Martens12}. The exact theory
has been used to map interacting bacterial baths into 
an effective non-interacting system \cite{Paoluzzi13}. 
Run-and-tumble is analytical tractable to study
sedimentation and harmonic trapping \cite{Tailleur09}, rectification \cite{Tailleur09,Angelani11b}, 
first-passage time problems in one dimension \cite{Angelani14}, self-trapping and collective phenomena \cite{Tailleur08}. 
By means of run-and-tumble model, ratchet phenomena can be studied analytically \cite{Tailleur09,Angelani11b} and through numerical simulations \cite{Angelani09,Angelani10,Angelani11}.

A central quantity in run-and-tumble dynamics is the persistence length $l=v_0 / \lambda$
that is fixed by both the tumbling rate $\lambda$,
and the self-propulsion velocity $v_0$.  The persistence length sets 
the crossover between a ballistic regime at short length scales and
a diffusive regime over long distances. The diffusive regime is characterised by a diffusivity  
$D=v_0^2/d \lambda$ \cite{Cates12} with $d$ the dimensionality of the space. 
Generalizing the Stokes-Einstein equation we can associate to run and tumble particles an effective thermal energy scale
defined by $D=\mu k_B T_{eff}$ where $\mu$ is the mobility. 
In many situations active particles have been actually found to behave like hot colloids \cite{Maggi13} with an effective temperature given by $T_{eff}$ \cite{Tailleur09}.
However, at variance with Brownian motion, where the thermal noise is practically unbounded, the propelling force in swimming bacteria has a finite value that sets the maximum slope that
bacteria can climb when escaping from an energy barrier.

We used numerical simulations to study the dynamics of run and tumble bacteria moving in the random energy landscape generated by the intensity of a speckle field. When increasing the overall intensity of speckles, we observe a crossover between a homogeneous and an inhomogeneous density regime where the density is enhanced
on the intensity maxima of the speckles. The crossover is characterised by a decrease in the configurational entropy
and by the emergence of a plateau in the collective part of the intermediate scattering function.
A similar crossover is expected to occur for Brownian particles when the average value of the random energy landscape increases above the thermal energy scale $k_B T$. Here we found that the crossover for active particles occurs before the average landscape energy reaches $k_B T_{eff}$. A much better estimate for the position of the crossover is obtained by equating the maximum external force to the propelling force of bacteria.

The paper is organized as follows.
In Sec. \ref{model} we introduce the model for the speckle field, in Sec. \ref{num} we illustrate
the numerical methods, in Sec. \ref{results} we present and discuss the results. 
\section{Speckle field in numerical simulations}\label{model}
\begin{figure*}[!t]
\centering
\includegraphics[width=1.\columnwidth]{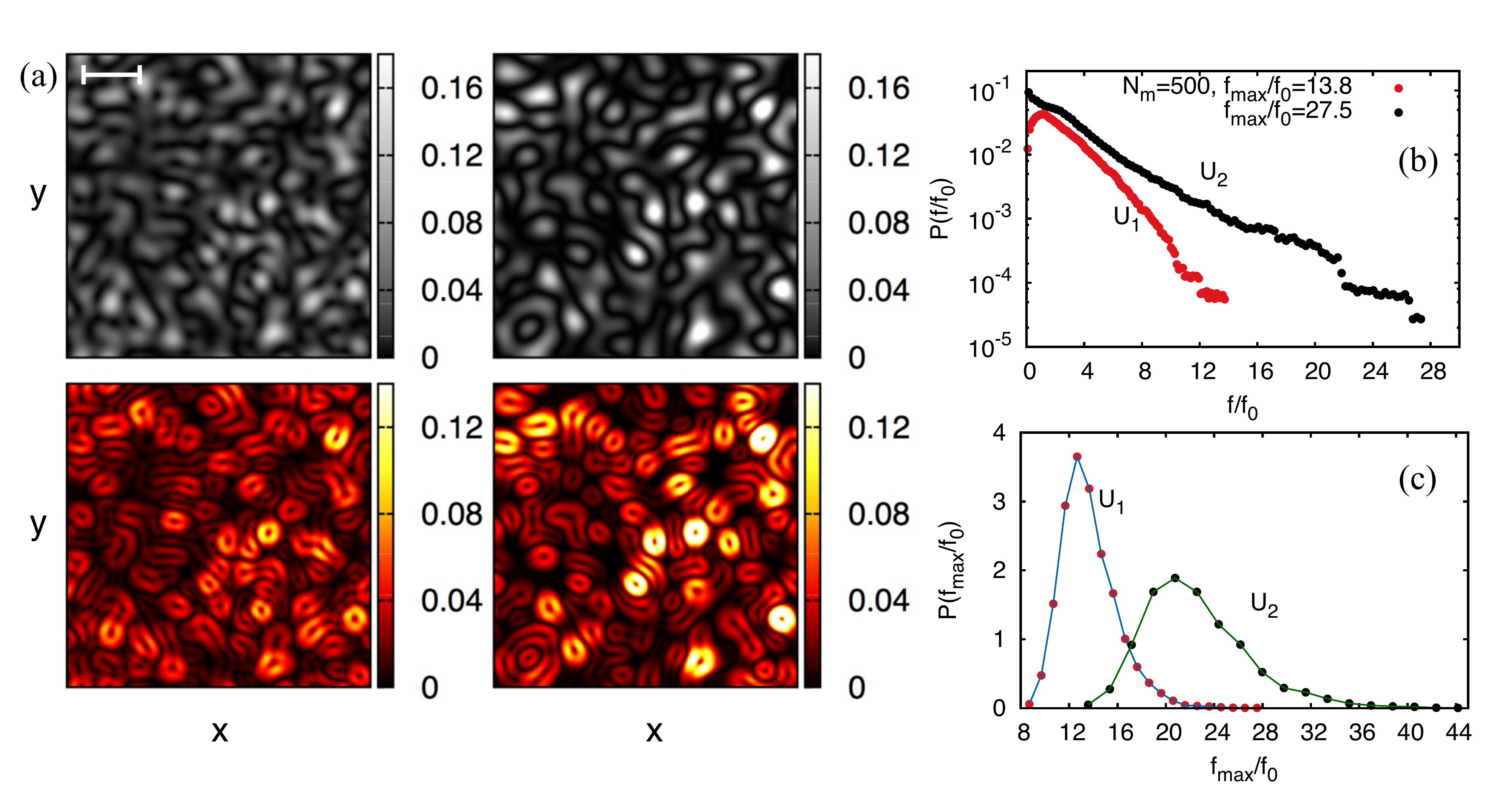}
\caption{{\bf{Speckle patterns and force distributions.}} Panel (a): in the top the two speckle patterns $U_1$ (left) and $U_2$ (right)
used in the simulations ($c=0.165$). The scale for the intensity is mW$\mu$m$^{-2}$; in the bottom we report the contour plot of the modulus of the force in pN. The white scale bar corresponds to $30\mu$m.
Panel (b): distribution of the force of the sample used in the numerical simulations. Panel (c): distribution of the maximum of the force obtained averaging over $N_s=3000$ samples.
The forces are expressed in unit of the self-propulsion of the swimmer $f_0$.}
\label{fig:varc}       
\end{figure*}

A speckle field can be obtained as the superposition of $N_m$ Fourier modes where both wave vectors and phases are randomly chosen:

\BEQ\label{field}
\varphi(\mathbf{r})=c \sqrt{\frac{k_B T_{eff} }{ N_m}}\sum_{l}  e^{i(\rr\cdot \mathbf{k}_l + \theta_l)} \, ,
\EEQ
%
with $\rr=(x,y)$.
Imposing periodic boundary conditions $\mathbf{k}_l=(2 \pi/L)(l_x\,\hat x +l_y\,\hat y)$, where $L$ is the box length and $l_x, l_y$ are random positive and negative integers satisfying the condition $|\mathbf k_l|<2\pi/\ell$ with $\ell$ the cell length.
The random phases $\theta_l$ are uniformly distributed between $0$ and $2\pi$.
The square modulus of $\varphi(\rr)$ is a real scalar function having the dimensions of energy and an exponential distribution \cite{goodman} with average $c^2 k_B T_{eff}$. The parameter $c$ is a dimensionless number that tunes the intensity of the forces.
As discussed in the following, we model steric interactions between elongated cell bodies using two force centers arranged along the cell axis. The mechanical action of speckles is consequently calculated as a system of forces acting on the same two centers and equal to the gradient of the speckles intensity:
\BEQ\label{spec}
I(\rr) = \left\{ \Re[\varphi(\mathbf{r})] \right\}^2 + \left\{ \Im[\varphi(\mathbf{r})] \right\}^2\, ,
\EEQ
where $\Re$ and $\Im$ are respectively the real and the imaginary part.
The forces due to the field attract bacteria towards region of high intensity
and can be expressed as
\BEQ\label{extf}
\mathbf{f}_{ext}(\rr)=-\nabla U(\rr) 
\EEQ
where the potential $U(\rr)$ is
\BEQ\label{rel}
U(\rr)=-I(\rr) \, .
\EEQ
From Eq. (\ref{rel}) follows that speckles play the role of a random energy landscape \cite{Dean07}.
The maximum value of the energy is zero, and the energy of the local minima, i. e.,  the light spot of the speckles, 
depends on the local intensity of the light.
In the following we will consider two cases. The first one is the usual speckle field obtained from the interference of the random Fourier modes
\BEQ\label{speck1}
U_1(\rr) = - \left\{ \Re[\varphi(\mathbf{r})] \right\}^2 - \left\{ \Im[\varphi(\mathbf{r})] \right\}^2\,  
\EEQ
The second case is obtained taking only the real part of $\varphi$:
\BEQ\label{speck2}
U_2(\rr) = - 2\left\{ \Re[\varphi(\mathbf{r})]\right\} ^2 \, 
\EEQ
giving rise to slight different spatial pattern and that can be
%
easily obtained in a laboratory with the aid of spatial light modulators.
The real and imaginary parts of $\varphi(\rr)$ vanish over independent curved paths on the $x,y$ plane \cite{goodman}.
This implies that while $U_1$ only vanishes at the intersection points of those paths, $U_2$ goes to zero over the entire length
of the paths $\Re[\varphi(\rr)]=0$. Therefore, when the average energy in the two potentials is equal, we expect to observe larger spatial gradients, and hence forces,  for $U_2$.
The speckle patterns used in the simulations are generated using the same realization of $N_m=500$ wave numbers and phases. 
The intensity of the speckles for $c=0.165$ are shown in the top of panel (a) of Fig. (\ref{fig:varc})
($U_1$ in the left of the panel (a) and $U_2$ in the right of the same panel). In the bottom of the panel (a)
we show the contour plot of the modulus of the force field.
The two patterns contain the same energy but, as we can see from the contour plot of the modulus of the forces,
the speckle $U_2$ exerts greater forces than $U_1$.  


In the panel (b) of Fig. (\ref{fig:varc}) we report the 
probability distribution of the force for the patterns 
used in the simulations. 
In unit of $f_0$, the self propulsion of the swimmer, setting $c=1.0$, the mean value is $2.4$ for $U_1$ and $3.0$ for $U_2$.
The maximum force is $13.8$ for $U_1$ and $27.5$ for $U_2$. Using the parameter $c$
to control the intensity of the speckle, the mean force acting on entire the swimmer equals the self propulsion when $c=0.460$
for $U_1$ and $c=0.410$ for $U_2$. Looking at the maximum value of the force, the self propulsion
is matched at $c=0.190$ for $U_1$ and $c=0.135$ for $U_2$.  

 In the panel (c) of Fig. (\ref{fig:varc}) we show the probability distribution of the maximum value of force for $U_1(\rr)$ and $U_2(\rr)$.
The figure is obtained averaging over $N_s=3000$ samples of
speckles with same energy.  
As expected, according to the shape of the distribution the speckle $U_2(\rr)$ is characterized by a long tail for large values of $f$.


\section{Molecular Dynamics Simulations}\label{num}
\subsection{Run-and-Tumble dynamics}
\begin{figure}[!t]
\centering
\includegraphics[width=0.45\columnwidth]{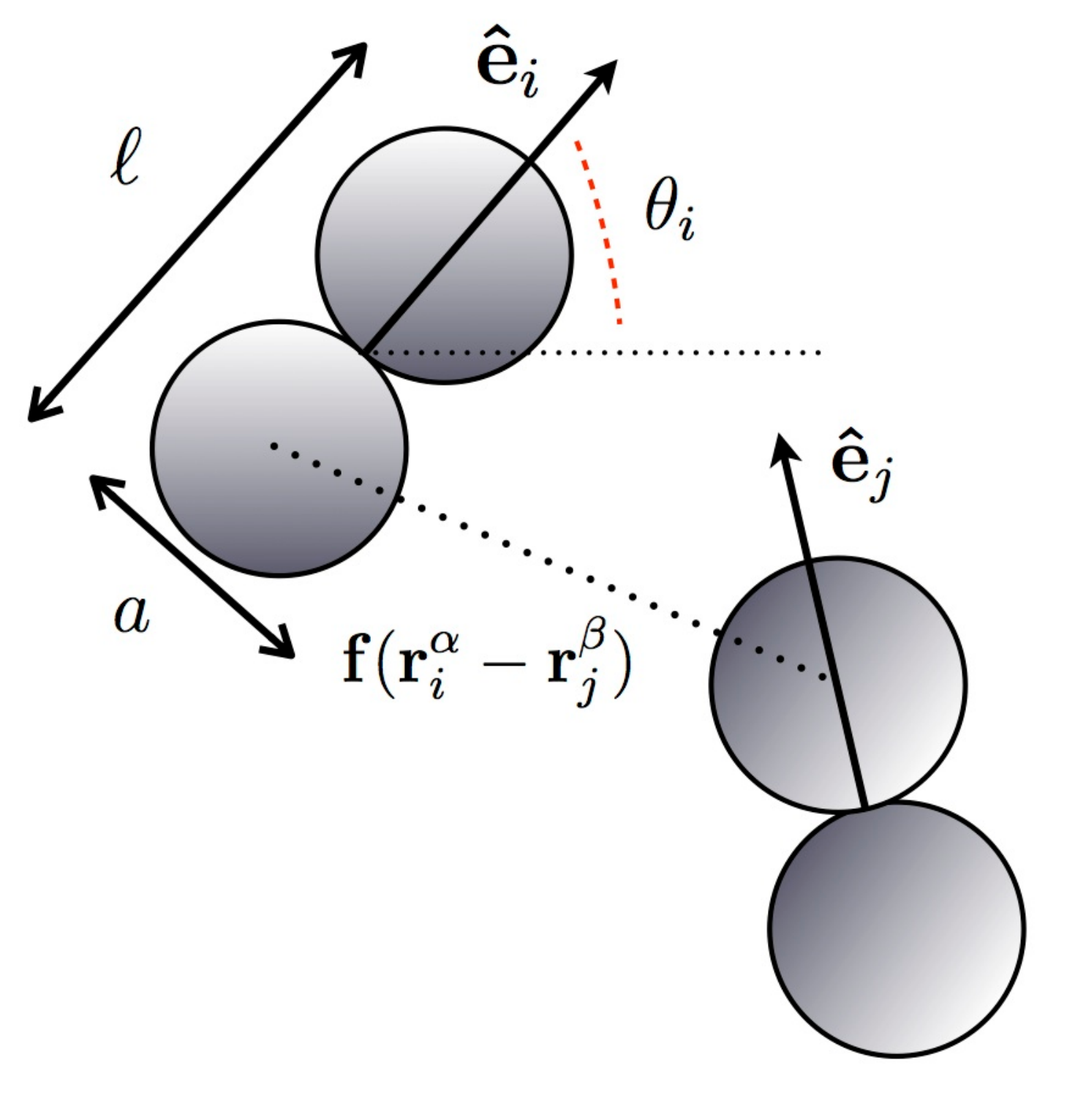}
\caption{{\bf{Pictorial representation of the model.}} Two swimmers, labelled by $i$
and $j$, are modeled by a unit vector $\mathbf{e}$ representing the swimming direction.
Along the swimming direction are located two force-centers (beads) labelled by greek symbols $\alpha=1,2$
(for the swimmer $i$) and $\beta=1,2$ for ($j$). The interaction between two beads
(of different swimmers) is short-ranged and repulsive.}
\label{fig:mod}       
\end{figure}

We perform numerical simulations of run-and-tumble
dynamics in two dimensions. 
Considering a system of $N$
self-propelled swimmers each of length $\ell$
and thickness $a$ (for a detailed discussion of the model see \cite{Angelani09,Angelani10,Angelani11}), the swimmer is modeled by a unit vector $\mathbf{e}_i$,
representing the swimming direction, and two short-range
repulsive force-centers (beads) arranged along it.
The position of the two beads of the $i-$th cell is labelled by greek
symbols (the swimmers are represented in Fig. (\ref{fig:mod})). 

At low Reynolds numbers regime \cite{Purcell77,Kim05},
the equations of motion are
\BEA\label{eqm}
\mathbf{v}_i &=& \mathbf{M}_i \, \mathbf{F}_i \\ \nn
\boldsymbol{\omega}_i &=&\mathbf{K}_i  \, \mathbf{T}_i 
\EEA
where $\mathbf{v}_i$ is the center of the mass velocity and $\boldsymbol{\omega}_i$ the angular velocity of the $i-$th swimmer. $\mathbf{M}_i$ and $\mathbf{K}_i$ are the translational and rotational mobility matrices 
\BEA
\mathbf{M}_i &=& m_{\parallel} \mathbf{\hat{e}}_i  \otimes \mathbf{\hat{e}}_i  + m_{\perp} \left( \mathbb{1} -  \mathbf{\hat{e}}_i  \otimes \mathbf{\hat{e}}_i \right)   \\ \nn
\mathbf{K}_i  &=& k_{\parallel} \mathbf{\hat{e}}_i   \otimes \mathbf{\hat{e}}_i  + k_{\perp} \left( \mathbb{1} - \mathbf{\hat{e}}_i    \otimes \mathbf{\hat{e}}_i \right) \, ,
\EEA
the symbol $\otimes$ is the dyadic product and $\mathbb{1}$ the identity matrix.
In Eq. (\ref{eqm}), $\mathbf{F}_i$ and $\mathbf{T}_i$ are the total force and the
total torque acting on the swimmer
\BEA\label{fandt}
 \mathbf{F}_i  &=& f_0 \mathbf{\hat{e}}_i(1 - \sigma_i) + \sum_{j\neq i, \alpha, \beta} \mathbf{f}(\mathbf{r}_i^\alpha - \mathbf{r}_j^\beta) + \sum_{\alpha} \mathbf{f}_{ext}(\rr_i^\alpha)\\ \nn
 \mathbf{T}_i  &=& \mathbf{t}_r \sigma_i + \mathbf{\hat{e}}_i \times \left( \sum_{j\neq i,\beta \alpha} \delta^\beta \mathbf{f}(\mathbf{r}_i^\alpha - \mathbf{r}_j^\beta) + \sum_{\alpha} \delta^\alpha \mathbf{f}_{ext}(\rr_i^\alpha)\right) \, .
\EEA
The index $j=1,\dots,N$ runs over swimmers, the indices $\alpha=1,2$ and $\beta=1,2$ run over beads, and
$\sigma_i$ is a state variable, $0$ for running swimmers and $1$ for tumbling ones.
The position of the beads of the $i$-th swimmers is 
\BEQ
\rr_i^\alpha=\rr_i + \delta^\alpha  \mathbf{\hat{e}}_i\, ,
\EEQ 
where
\BEQ
\delta^\alpha=\left( 2 \alpha - 3\right)\frac{\ell}{4} 
\EEQ 
giving rise to, i. e.,  $\delta^{1,2}=\pm\frac{\ell}{4}$.
The pair force $\mathbf{f}(\mathbf{r})$ is the repulsive interaction among the swimmers (steric term)
\BEQ
\mathbf{f}(\mathbf{r})=\frac{A \mathbf{r}}{\mathbf{r}^{n+2}}\,, \,
\EEQ
where the coefficient $A$ is fixed by imposing that two swimmers facing
head to head on the same line would be in equilibrium at the distance $a$
\BEQ
A = f_0 a^{n+1}\, 
\EEQ
where we choose $n=12$. In Eq. (\ref{fandt}) the external force $f_{ext}$ is given by expression (\ref{extf}) and $f_0$ is the self-propulsion force.
The two-state variables $\sigma_i$ stochastically change
with rate $\lambda$ from the value $1$ (tumbling state) to $0$ (running state).
In the tumbling state the $i-$th cell changes the free swimming direction 
due to a random torque $\mathbf{t}_r$ acting for a finite tumbling-time $\tau=\lambda^{-1}/10$ (this value of $\tau$ is suitable for {\itshape E. coli} cells \cite{Berg04}), in the following we consider $a=\ell/2$.
The system is enclosed in a square box of side $L$ with
periodic boundary conditions.

The equations of motion are numerically integrated for $T=100$s by means of a second-order Runge-Kutta scheme
with a time step $\Delta t=10^{-4}$s. Choosing realistic parameters for {\itshape E. coli}, we have $\ell= 3\, \mu$m, 
$m_\parallel = 60\, \mu$m s$^{-1}$pN$^{-1}$, $f_0= 0.5$ pN, ($v_0=30 \mu$m s$^{-1}$),
$\lambda^{-1}=1$ s and $\tau=0.1$ s.
To study the Brownian limit we change the time steps from $\Delta t=10^{-4}$s to $\Delta t=10^{-5}$s.
The mobility parameters are chosen as $k_{\perp}=31.3\, \mu$m s$^{-1}$, $m_\parallel= 60\, \mu$m s$^{-1}$pN$^{-1}$ and $m_\perp=52.2\, \mu$m s$^{-1}$pN$^{-1}$ \cite{Angelani09}. 
The relations between physical and internal units are reassumed in Tab. (\ref{tab:internal}).
Performing two dimensional simulations, $k_{\parallel}$ does not
play any role. 
We investigate
non interacting and interacting swimmers at density, $\bar{\rho}=N/L^2=0.018,0.028\,\mu$m$^{-2}$ at fixed $L=150\mu$m.  
The non-interacting case, i. e., a gas of run-and-tumbe particles, is obtained switching off the steric potential.
The Boltzmann limit is studied increasing tumbling rate $\lambda^{-1}=1.0,0.25,0.1,0.025,0.01$s and free swimming velocity $v=30,60,94.9,189.7,300\, \mu$ms$^{-1}$. 
The two fields $U_1(\rr)$ and $U_2(\rr)$ are generated 
by the same realization of $N_m=500$ wave vectors $\mathbf{k}_l$ and phases $\theta_l$.
The field is evaluated on a grid of $10^5 \times 10^5$ points by means of parallel OpenMP algorithm.

\begin{table}
\centering
\resizebox{1.00\columnwidth}{!}{
\begin{tabular}{||         c               |     c                   |    c                      |         c          |    c             |    c                         |     c          |   c                                     |     c           |    c                                                        |   c                        |   c    |  c ||}  \hline 
                            &    $\ell$           &  $a$               & $L$                  & $\Delta t$ & $T$          & $\lambda^{-1}$ & $\tau $ & $v_0$                                 & $f_0$      &  $m_\parallel$                                 &  $m_\perp$      & $k_\perp$   \\ \hline
Internal Units    &    $1$              &  $1 / 2$          & $50$               & $10^{-3}$ & $10^3$    & 	$10$	                &	   $1$       & $1$                                & $1$	     &         $1$                                            &     $0.87$           &  $4.8$  \\ 
Physical Units  &    $3 \mu$m   &  $1.5\mu$m  & $150 \mu$m  & $10^{-4}$s       & $100$s    & $1$s		       &   $0.1$s  &$30\mu$m s$^{-1}$    & $0.5$pN &  $60\, \mu$m s$^{-1}$pN$^{-1}$ &  $52.2\, \mu$m s$^{-1}$pN$^{-1}$ & $31.3\, \mu$m s$^{-1}$         \\
\hline
\end{tabular}}
\caption{{\bf{Internal and physical units.}} Values (in internal and physical units) of the parameters used in the simulations.}\label{tab:internal}
\end{table}

\subsection{Methods}\label{methods}

For a given realization of the speckle fields, varying the intensity of the external forces through the parameter $c$, 
we investigate the ergodicity of the system looking at the behavior of dynamic observables, e. g., the correlation of the density fluctuations,
and static observables, e. g., the density profiles and the probability distribution of the velocity.

The correlation of the density fluctuations is given by the intermediate
scattering function. We compute both, the collective $F_{coll}(\mathbf{q},t)$
and the self $F_{self}(\mathbf{q},t)$ intermediate scattering function
\BEA
F_{coll}(\mathbf{q},t)&=&\frac{1}{N} \left \langle \sum_{l,m} \exp{\left[ {-i \Delta \rr_{lm}(t,t^\prime)\ \cdot \mathbf{q} }\right] } \right\rangle_{t^\prime}  \\ \nn
F_{self}(\mathbf{q},t)&=&\frac{1}{N} \left \langle \sum_{l} \exp{\left[ {-i \Delta \rr_{ll}(t,t^\prime)\ \cdot \mathbf{q} }\right] } \right\rangle_{t^\prime} 
\EEA
with 
\BEQ
\Delta \rr_{lm}(t,t^\prime)\equiv\rr_l(t+t^\prime ) - \rr_m(t^\prime )
\EEQ
The averaging is defined as follows
\BEQ
\left\langle  \mathcal{O}(t) \right\rangle_{t} \equiv \frac{1}{T}\int_{t_0}^{T+t_0} \,dt\, \mathcal{O}(t)
\EEQ
and the initial time $t_0$ is chosen such that $t_0 > \lambda^{-1}$.
In our simulations we take $t_0=5\,$s.

\begin{figure}[!t]
\centering
\includegraphics[width=0.48\columnwidth]{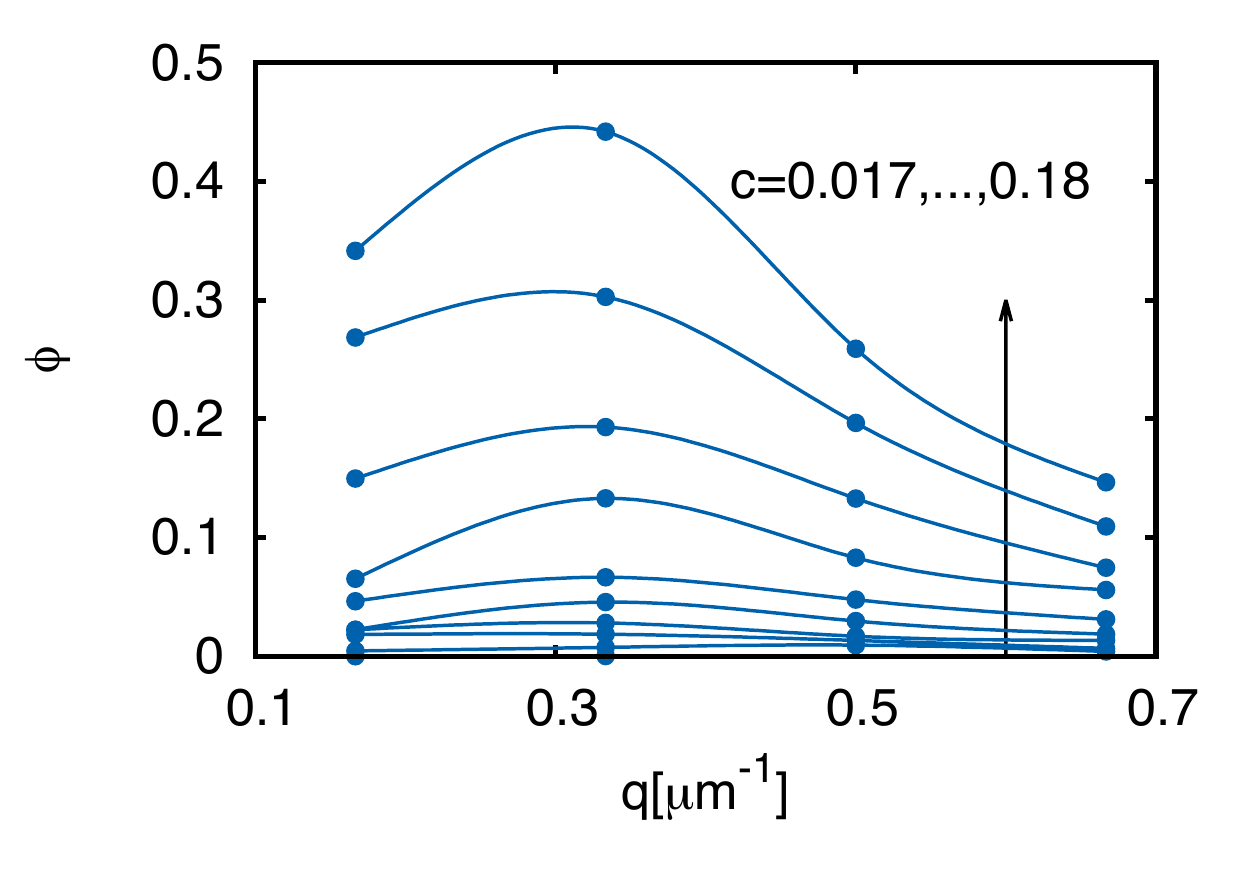}
\includegraphics[width=0.48\columnwidth]{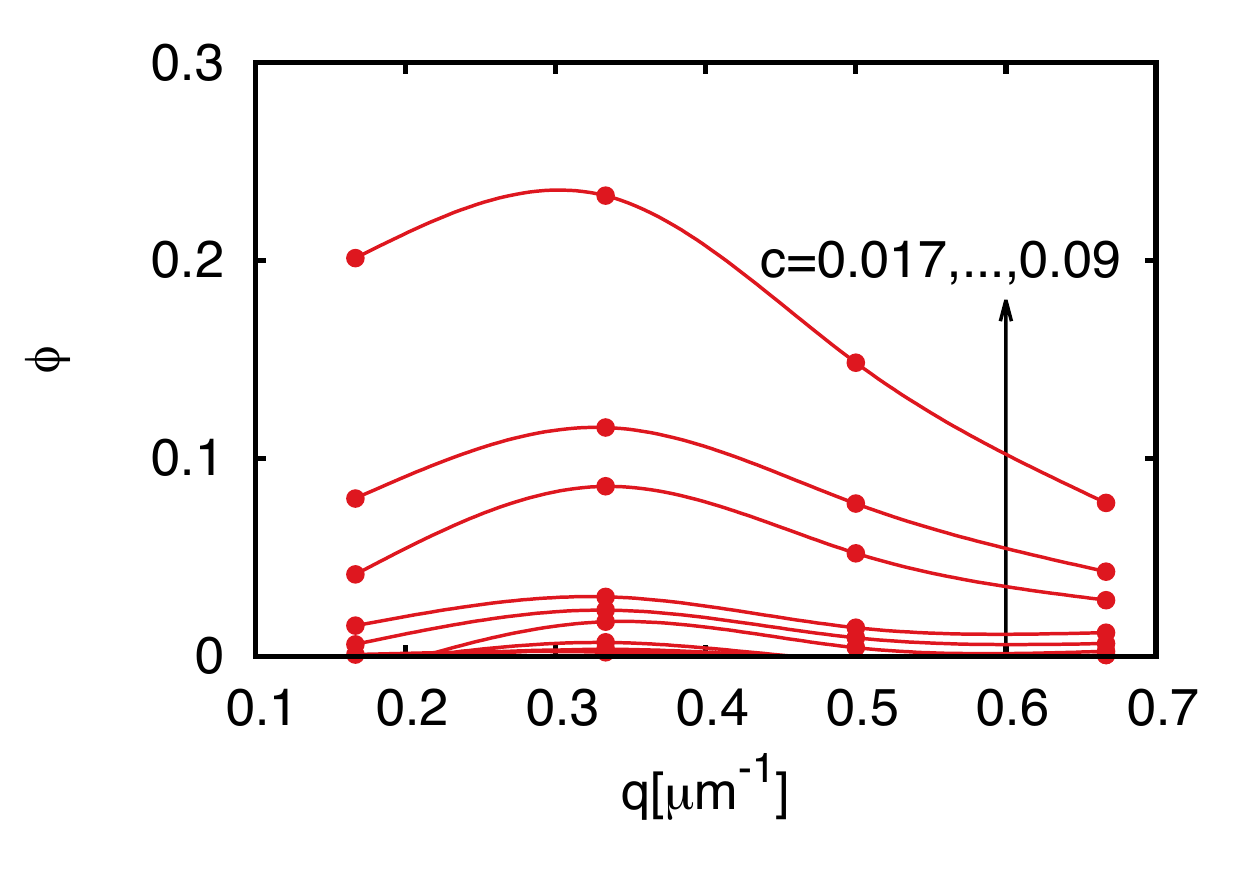}
\caption{{\bf{The ergodicity parameter at different wavenumbers.}}
Ergodicity parameter $\phi$ as a function
of $q$ for the pattern $U_1$ (left panel) and $U_2$ (right panel) at different values of $c$. 
Continuous lines are obtained by spline interpolation.}
\label{fig:phiq}       
\end{figure}

Varying the intensity of the external field, a finite number of 
swimmers spend more and more time in the minima of the random-energy landscape.
Looking at the long-time behavior of $F_{coll}(\mathbf{q},t)$ 
we define the ergodicity parameter as follows \cite{Gotze91}
\BEQ
\phi(c,q) \equiv \lim_{t\to\infty}F_{coll}(\mathbf{q},t) \, .
\EEQ
The ergodicity parameter gives a measure of the fraction of swimmers localized on the spatial scale $R_b\sim 1/q$. 
In Fig. (\ref{fig:phiq}) we show the dependency of $\phi$ on $q$ for $U_1$ (left panel)
and $U_2$ (right panel). 
The peak developed by $\phi$ at $q\sim0.3\,\mu$m$^{-1}$
signals the spatial scale of the regions of maximum speckle intensity.

To study the static properties of the model we start from
the density profile defined as
\BEQ
\rho(\rr)=\frac{1}{\mathcal{V}}\left\langle \sum_i \delta( \rr - \rr_i(t))\right\rangle_t \, ,
\EEQ
the normalization factor $\mathcal{V}$ is fixed by the condition 
\BEQ
\int d \rr \, \rho(\rr) = 1 \, .
\EEQ
The entropy of the distribution $\rho(\rr)$ reads
\BEQ\label{entropy}
s[\rho]=-\int d\rr\, \rho(\rr) \log{\rho(\rr)}
\EEQ
which, for $c\to0$, reduces to
\BEQ
\lim_{c\to0}s[\rho]=\log{V}
\EEQ
with $V=L^2$. 

Steady states for run-and-tumble particles are in general non-Boltzmann. 
The Boltzmann case can be obtained in the limit $\lambda,v_0\to\infty$ with
constant $v_0^2/2 \lambda$.
In the Boltzmann limit, the density profile in presence of the speckle field
$U_i(\rr)$, ignoring the excluded volume interaction, is 
\BEQ\label{boltr}
\rho_{B}(\rr)=\frac{e^{-\frac{U_i(\rr)}{k_B T_{eff}} }}{Z}
\EEQ
 with $i=1,2$. The normalization is
 \BEQ
Z=\int\, d\rr\, e^{-\frac{U_i(\rr)}{k_B T_{eff}}}
 \EEQ
The entropy of the distribution is
\BEQ\label{bolts}
s[\rho_B] = -\int\, d\rr\,  \rho_{B}(\rr) \log \rho_{B}(\rr)
\EEQ
and in the limit $c\to0$ one has 
\BEQ
\lim_{c\to0} s[\rho_B] = \log{V}=\lim_{c\to0}s[\rho]\, .
\EEQ
We compare the equilibrium properties of the isodiffusive simulations
with the Boltzmann limit expressed by Eq. (\ref{boltr}) with $k_B T_{eff}=D / \mu$.

Another static observable which gives information
about the density inhomogeneities, is the probability distribution of velocity 
for cells in the running state ($\sigma_i=0$)
\BEQ\label{pdv}
P(v)=\frac{1}{\mathcal{N}}\left\langle \sum_i \delta\left( v-v_i(t)\right | \sigma_i = 0) \right\rangle_t
\EEQ
with $v=|\mathbf{v}|$, $v_i(t)=|\mathbf{v}_i(t)|$ and $\mathcal{N}$ fixed by the condition
\BEQ
\int dv\,  P(v) = 1\, .
\EEQ
From the behavior of $P(v)$ at small $v$ we obtain information about the fraction
of particles locked by the field.

Finally, to study the transport properties of the system,
we look at the mean-square displacement
\BEQ\label{msd}
msd=\frac{1}{N} \left\langle \sum_i \left[ \rr_i(t+t^\prime) - \rr_i(t^\prime) \right]^2  \right\rangle_{t^\prime} \, .
\EEQ

\section{Results}\label{results}

\begin{figure*}[!t]
\centering
\includegraphics[width=0.5\columnwidth]{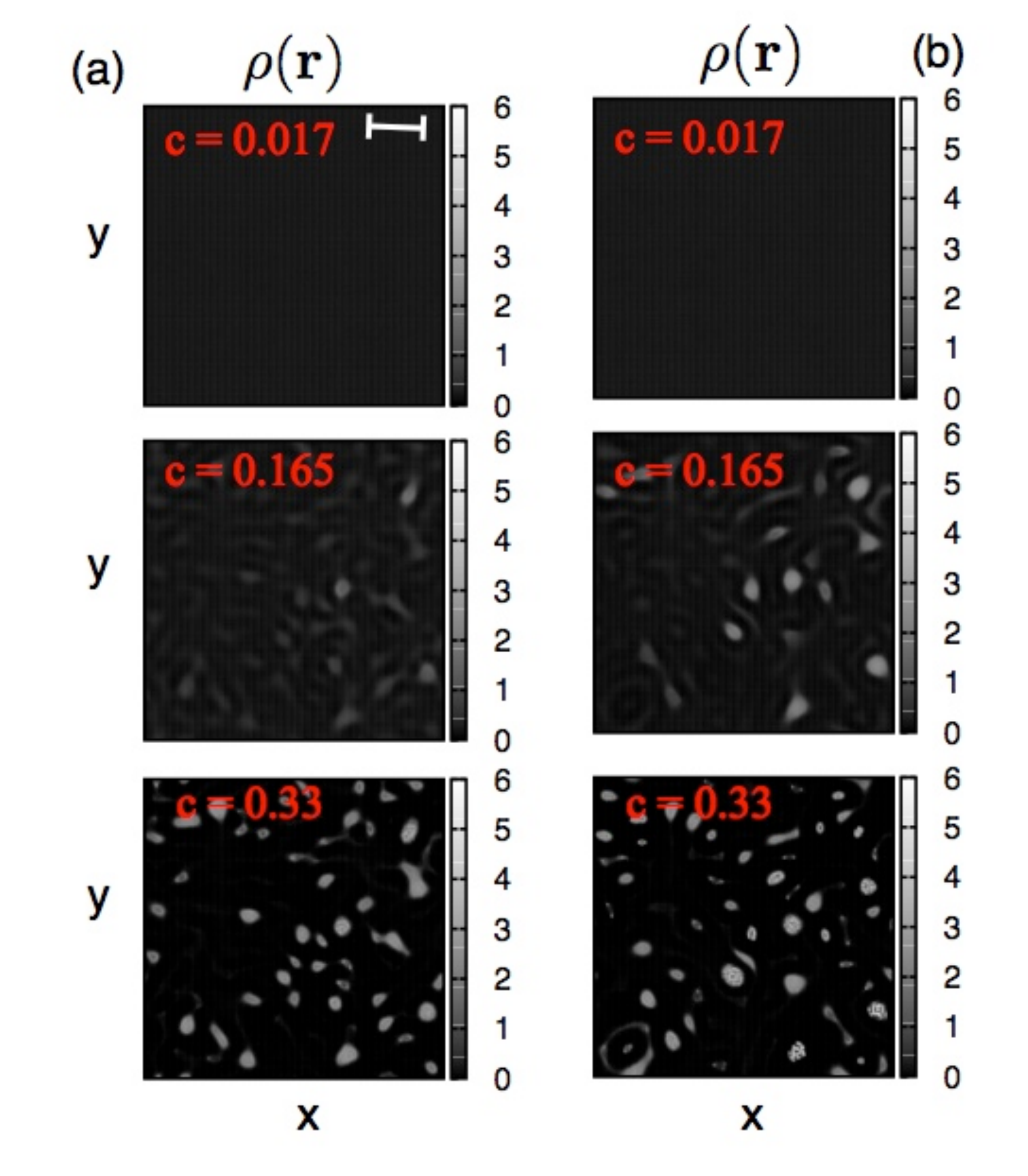}
\caption{{\bf{Density profiles at different speckle intensities.}}
Comparison between density profiles for $U_1(\rr)$ (panel (a)) and $U_2(\rr)$ (panel (b)) for
three values of the control parameter $c$.
The density field $\rho(\rr)$ is modulated by the random energy
landscape, increasing the intensity of the external field the
system spends more and more time in the minima of the
external potential breaking the ergodicity at high $c$. The white scale bar corresponds to $30\mu$m.} 
\label{fig:intrho}       
\end{figure*}

The speckles concentrates
$\rho(\rr)$ in the minima of the random energy landscape.
In Fig. (\ref{fig:intrho}) we show the density fields
at different values of the control parameter $c$ in the cases of speckle fields $U_1(\rr)$ (left panel) and $U_2(\rr)$ (right panel).
For the data shown in figure, the average bacterial density is $\bar{\rho}=0.018\,\mu$m$^{-2}$.
In the homogeneous phase ($c=0.017$ in Fig. (\ref{fig:intrho})),
the system is ergodic and the density is uniform in space. Increasing the 
intensity of the external field, density profiles become inhomogeneous 
and the bright spots in Fig. (\ref{fig:intrho}) indicate that the system spends more
and more time in the minima of the random energy landscape.

%
%
\begin{figure*}[!th]
\centering
\includegraphics[width=.8\columnwidth]{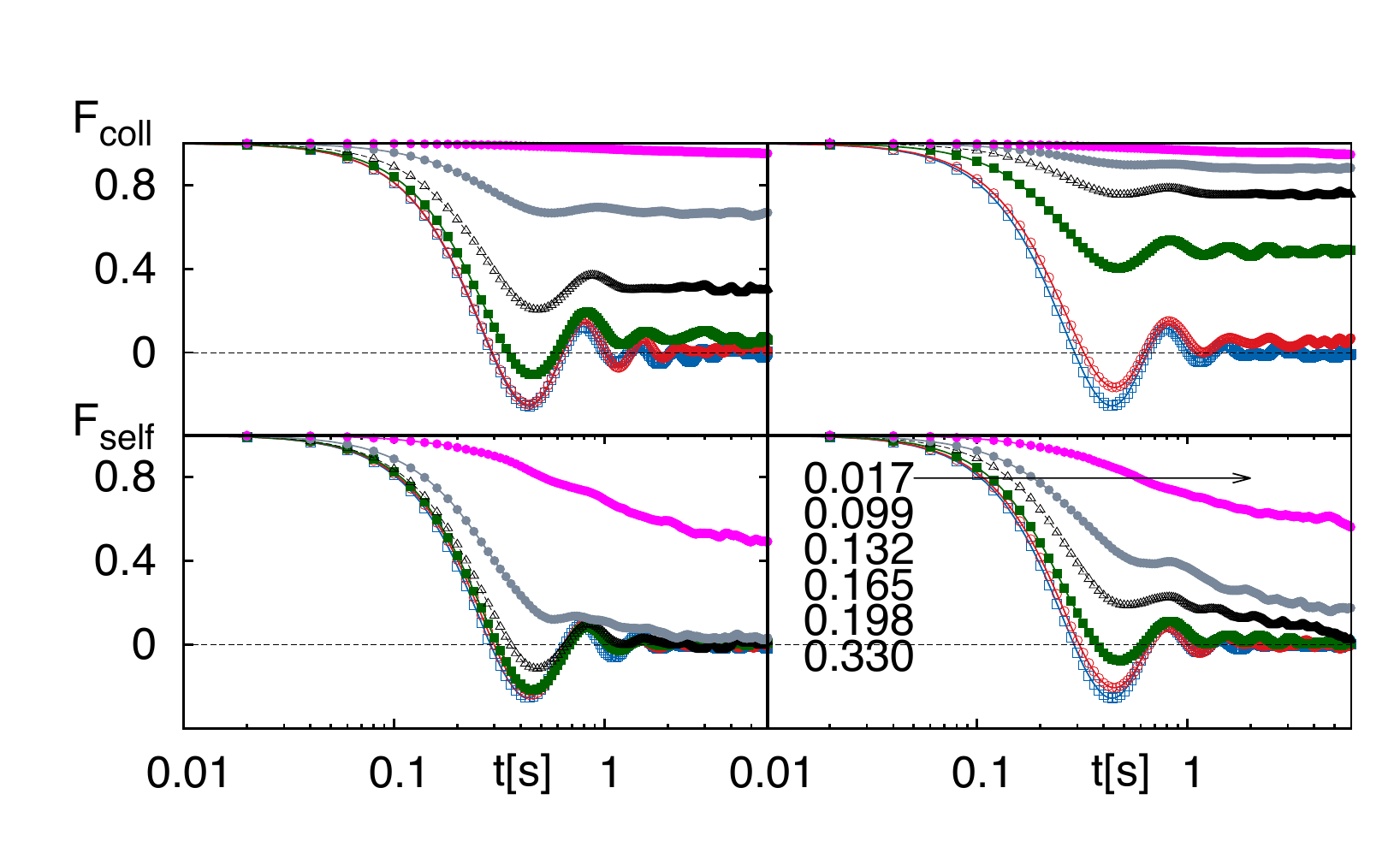}
\caption{{\bf{Intermediate Scattering Functions.}} Collective and self part of the Intermediate Scattering
function for $U_1(\rr)$ (left panels) and $U_2(\rr)$ (right panels)
for $q=0.3\,\mu$m$^{-1}$ at different values of $c$ from $c=0.017$ (blue) to $0.330$ (magenta).}
\label{fig:fcofself}       
\end{figure*}

\subsection{Threshold estimation}
\begin{figure*}[!t]
\centering
\includegraphics[width=0.5\columnwidth]{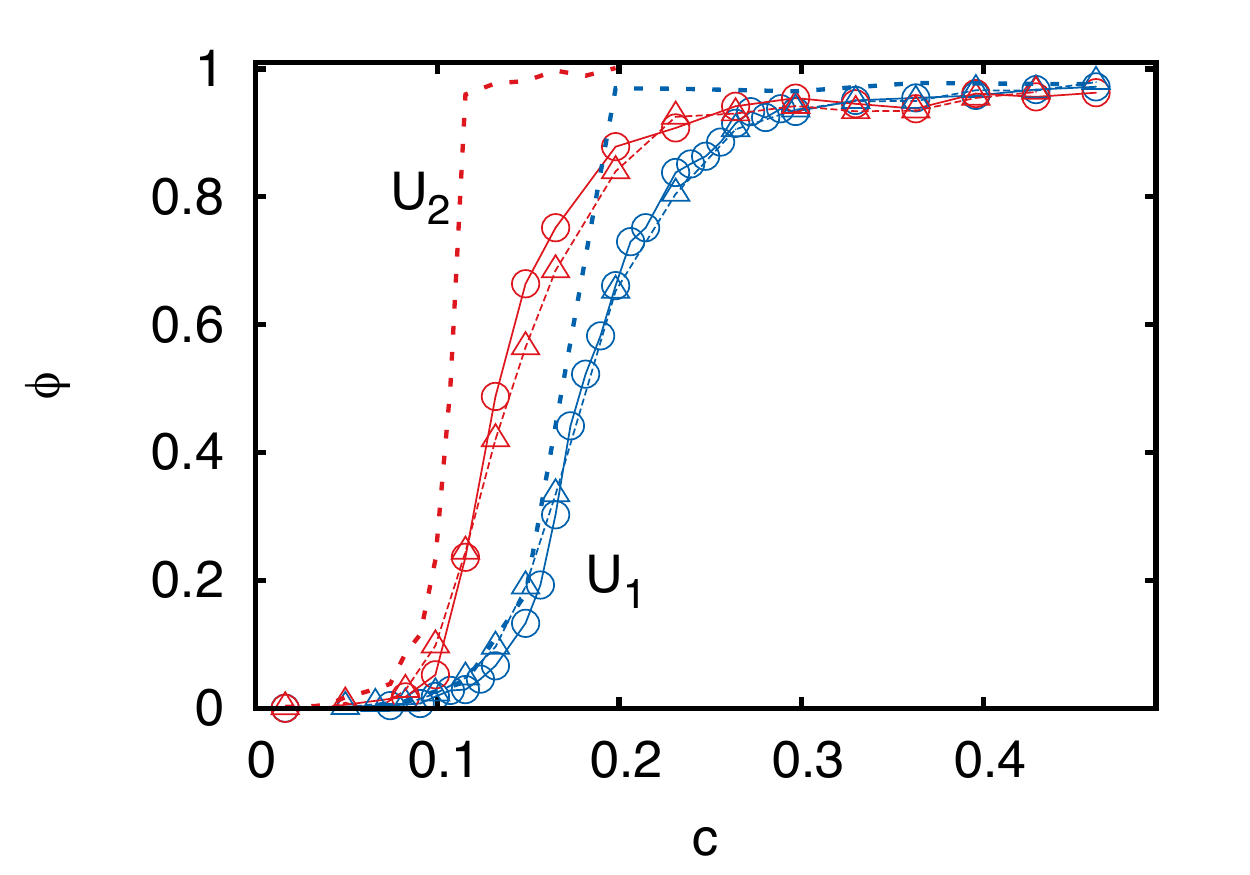}
\caption{{\bf{Ergodicity parameter as a function of the speckle intensity.}} The dashed lines
are the non-interacting (ideal gas of run-and-tumble bacteria) simulations, the circles 
represent simulations performed at density $\bar{\rho}=0.018\,\mu$m$^{-2}$ (triangle $\bar{\rho}=0.028\,\mu$m$^{-2}$). 
} 
\label{fig:plat}       
\end{figure*}

In Fig. (\ref{fig:fcofself}) we report the 
collective (top of the figure) and self (bottom) part of the intermediate
scattering function as a function of time at different intensities
of the speckle in the case of $N=400$ swimmers ($\bar{\rho}=0.018\,\mu$m$^{-2}$) and $q=0.3\,\mu$m$^{-1}$. 
Starting from $c\sim0.132$ for the potential $U_1$
($\sim0.099$ for $U_2$), the collective part of the intermediate scattering
function develops a plateau 
that continuously increases
from $0$. 
Looking at the self correlation,
in the range of $c$ where one has $\phi\neq 0$, 
$F_{self}(q,t)$ decays to zero, indicating that the single bacterium escapes from the energy barriers. 
At high enough values of $c$, $F_{self}(q,t)$ too does not decay to zero and a finite fraction of bacteria 
are trapped.
The threshold value $c^*_\phi$ has been defined looking at the maximum of $d\phi/dc$.
The behavior of $\phi$ as a function of $c$ is shown in Fig. (\ref{fig:plat}) for $\bar{\rho}=0.018,0.028\,\mu$m$^{-2}$ and for
non-interacting swimmers.
Looking at the interacting case, one has that the density does not play a crucial role on the trapping.
Comparing the interacting bacteria with the non-interacting ones, we observe that the excluded volume smooths the transition.
We obtain $c^*_\phi=0.165$ for $U_1$ and $c^*_\phi=0.116$ for $U_2$.

\begin{figure}[!t]
\centering
\includegraphics[width=0.5\columnwidth]{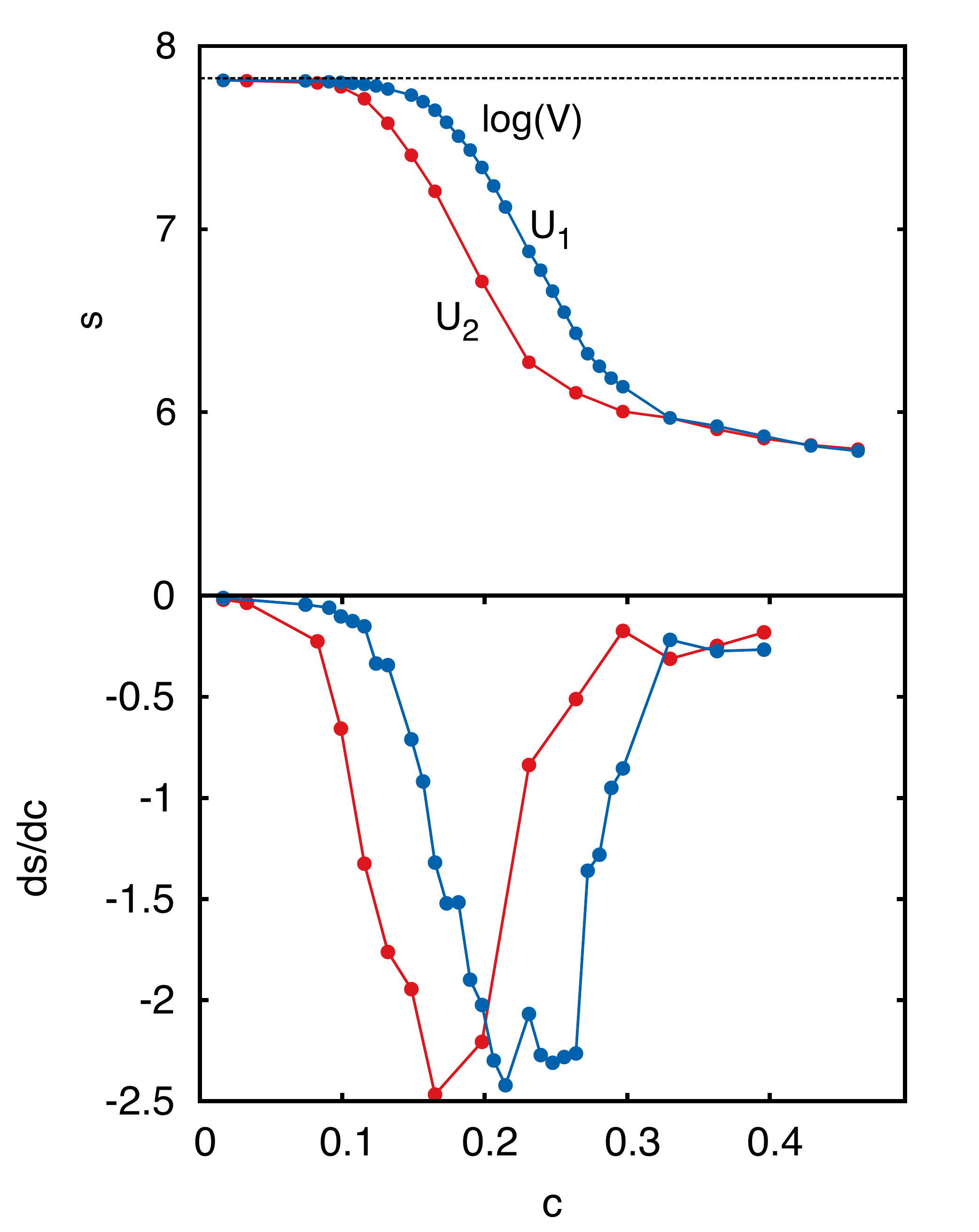}
\caption{{\bf{Entropy as a function of the speckle intensity.}} Top panel: the entropy
computed using Eq. (\ref{entropy}) varying the intensity of the speckle patterns. Bottom panel: derivative
of the entropy with respect the control parameter $c$. 
} 
\label{fig:entro}       
\end{figure}

Looking at the entropy defined by Eq. (\ref{entropy}),
we can give another estimation of the threshold value $c^*$. 
In Fig. (\ref{fig:entro}), we show the entropy as a function of $c$
for the speckle patterns $U_1$ and $U_2$ (top of the figure). 
The derivative of the entropy with respect to $c$ is shown  in the bottom of Fig. (\ref{fig:entro}).
The crossover value, defined as the minimum of $ds/dc$,
is $c^*_s=0.215$ for $U_1$ and $0.165$ for $U_2$.

\begin{figure*}[!th]
\centering
\includegraphics[width=0.45\columnwidth]{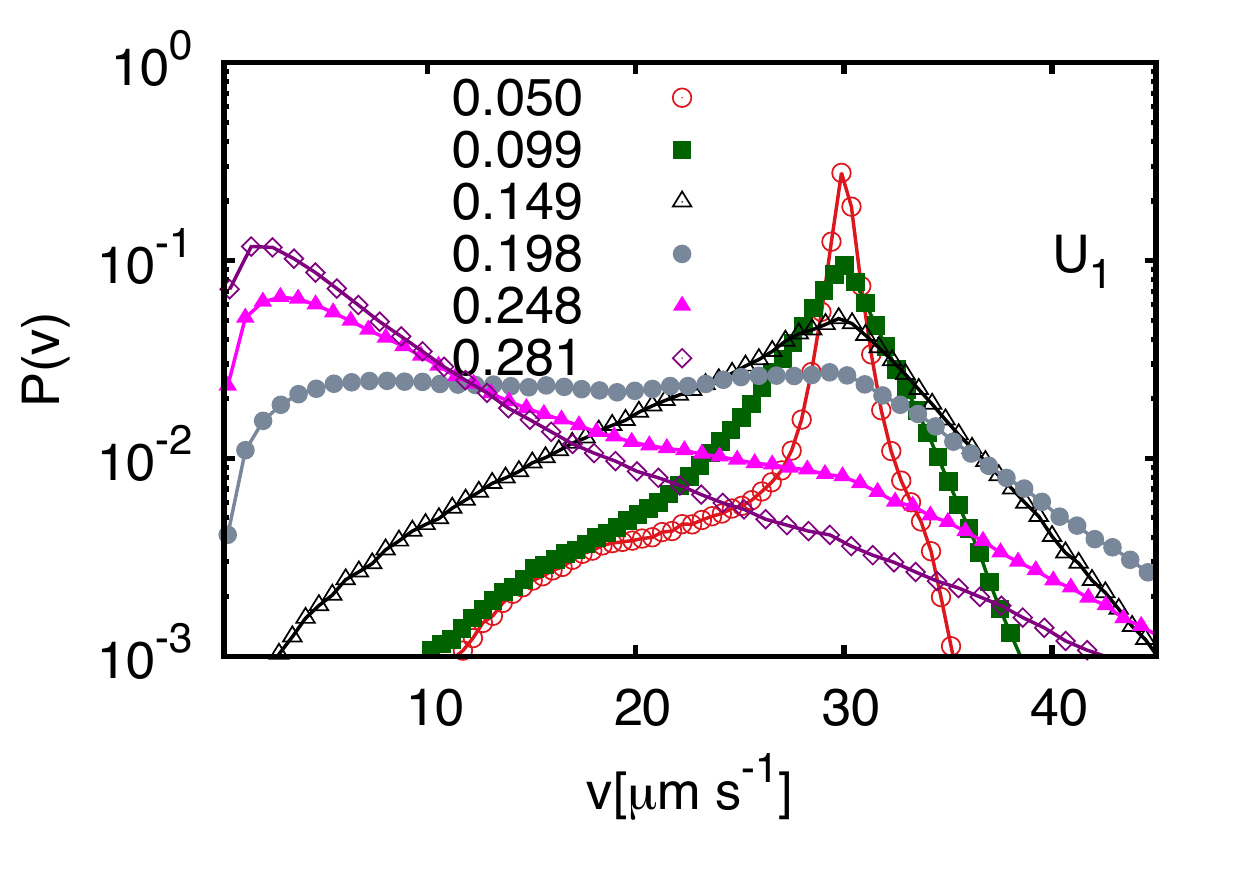}
\includegraphics[width=0.45\columnwidth]{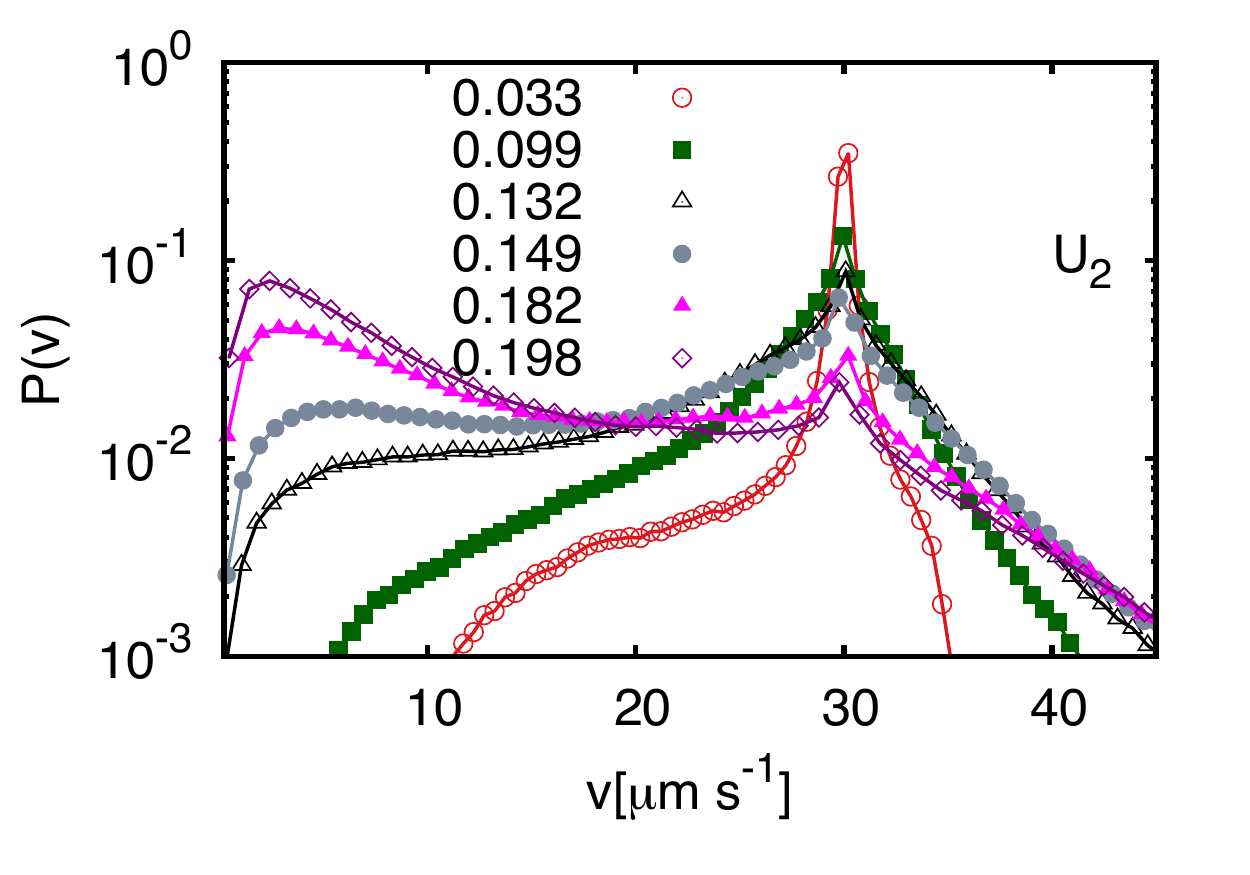}
\caption{{\bf{Probability distribution of velocity.}} Probability distribution of velocity for 
particles in the run-state in the presence of potentials $U_1(\rr)$ ($U_2(\rr)$, right panel) at different $c$.}
\label{fig:vel}       
\end{figure*}

A qualitative estimation of crossover value $c^*$ can be also obtained from the
probability distribution function of velocity defined by Eq. (\ref{pdv}).
Since $P(v)$ is computed by considering only particles in running state,
the peak at low velocities is due to the fraction of trapped particles
and the height is proportional to the number
of particles in the minima of the potential. 
In Fig. (\ref{fig:vel}) we report $P(v)$ for $U_1$ (left panel) and $U_2$ (right panel).
Increasing the intensity of the external field, the probability distribution of velocity
shows two peaks due to the competition between self-propulsion and trapping. 
The peak at high velocity is due to the self-propulsion and it is less pronounced
for the pattern $U_1$.

We can heuristically define $c^*_v$ as the value of $c$ for which
$P(v)$ becomes flat at low $v$. We have $c^*_v\sim 0.198$ for $U_1$ and $\sim 0.165$
for $U_2$. The threshold values are summarized in Tab. (\ref{tab:thr}).
The last column reports $c_{max}$,
defined as the value of $c$
where the maximum force of the speckle $f_{max}$ is equal to
the self-propulsion of the swimmer.
As one can see, all the values $c^*$ are of the same order
of magnitude of $c_{max}$.
\begin{table}
\centering
\resizebox{.4\columnwidth}{!}{
\begin{tabular}{|| c  | c  | c | c | c ||} \hline 
             & $c^*_s$  & $c^*_\phi$ & $c^*_v$ &$c_{max}$          \\ \hline
$U_1$ &    $0.215$   &  $0.165$         & $0.198$      &$0.190$                   \\ 
$U_2$ &    $0.165$   &  $0.116$         & $0.165$      &$0.135$                  \\
\hline
\end{tabular}}
\caption{Threshold values for the intensity of the speckles $U_1$ and $U_2$.
Estimation through the derivative of the entropy ($c^*_s$),
the derivative of the ergodicity parameter ($c^*_\phi$), looking
at the probability distribution of the velocity ($c^*_v$) and by
the maximum force of the speckle pattern.}\label{tab:thr}
\end{table}
Comparing the crossover values of $c$ for $U_1$ and $U_2$ we
can conclude that the pattern generated by $U_2$ spends less energy than $U_1$ to trap.
This is due to the tails in the distribution of the forces shown in Fig. (\ref{fig:varc}) and it is in agreement with the statistical properties of the distribution of the maximum force reported in Sec. (\ref{model}).

%

\subsection{Comparison with the Boltzmann limit}

In this section we investigate the relation between run-and-tumble dynamics on long time and Boltzmann equilibrium.
Run-and-tumble dynamics is diffusive on long time, on the other hand at the equilibrium the probability
distribution becomes Boltzmann only in limit cases \cite{Tailleur09,Cates12}.
The steric interaction changes the value of the diffusivity
from $D$ to $D_{int}$ with $D_{int}\leq D$ \cite{Paoluzzi13} 
\begin{figure}[!th]
\centering
\includegraphics[width=0.45\columnwidth]{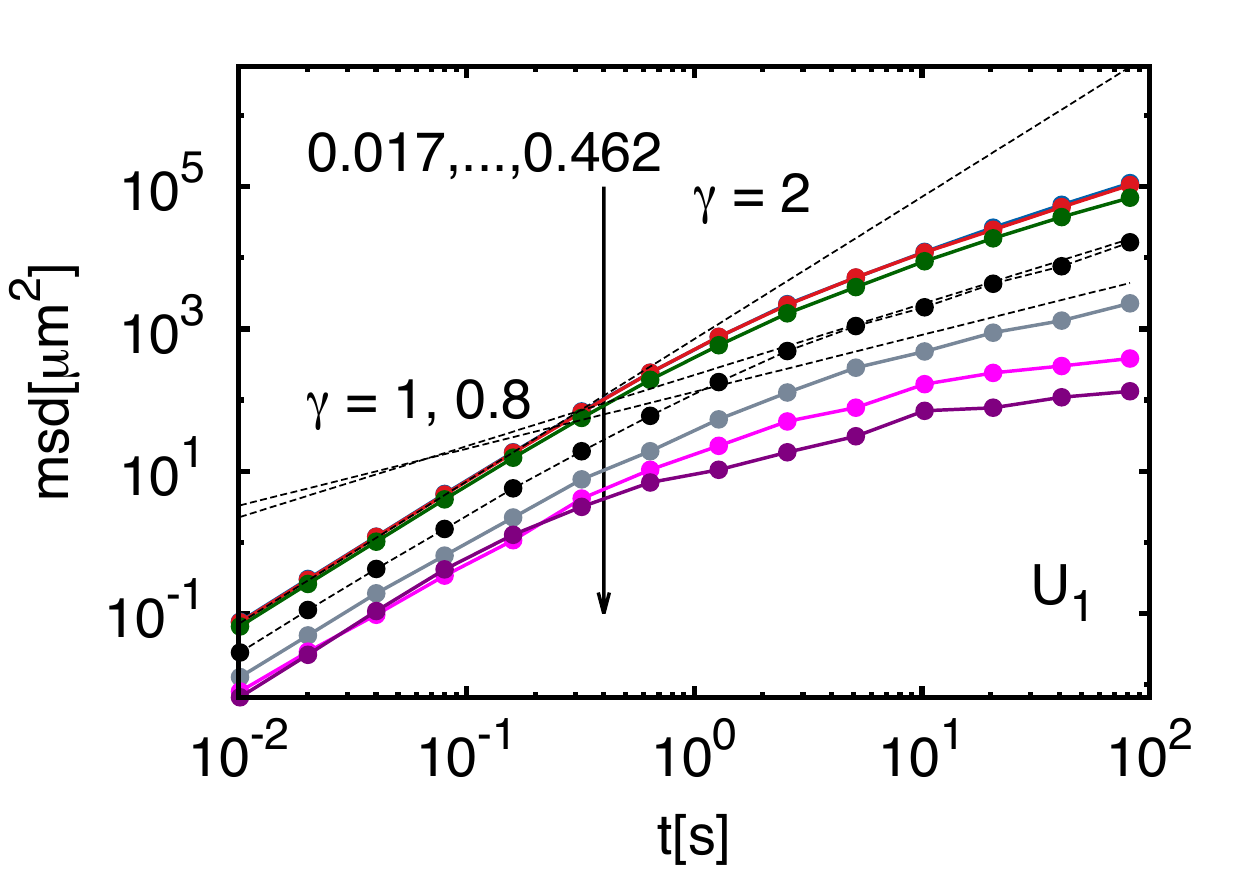}
\includegraphics[width=0.45\columnwidth]{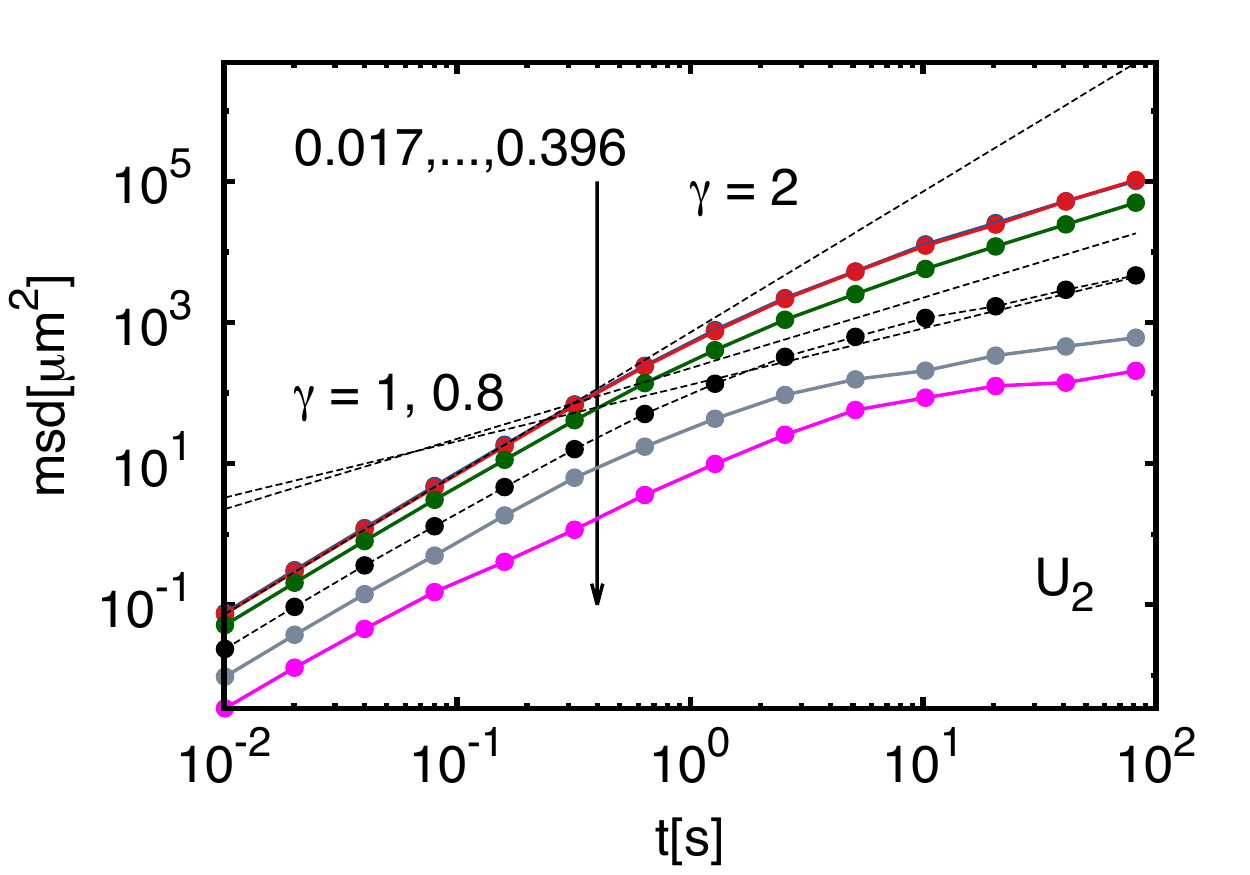}
\caption{{\bf{Subdiffusive regime at high intensities.}} Mean-square displacement for $U_1(\rr)$ ($U_2(\rr)$, right panel) 
at different $c=0.017,0.083,0.165,0.248,0.330,0.396,0.462$ for $U_1$ and
$c=0.017,0.083,0.165,0.248,0.330,0.396$ for $U_2$.
The dashed lines show the crossover between ballistic and diffusive regime. At high
$c$ the diffusive regime becomes subdiffusive.}
\label{fig:msd}       
\end{figure}
%
%
%
that can be obtained by the mean-square
displacement given by Eq. (\ref{msd}). In the long-time
limit one has:
\BEQ
msd\sim D_{int} \, t^\gamma \, .
\EEQ
In Fig. (\ref{fig:msd})
we show the mean-square displacement at different $c$
for $U_1$ and $U_2$. We observe normal diffusion ($\gamma=1$)
at small speckle intensity, and subdiffusion ($\gamma<1$) at higher intensities.
It is known in literature that Brownian particles 
embedded into random energy landscape show a subdiffusive
regime \cite{Bouchaud90,Zwanzig88,Novikov11,Hanes12a,Hanes12b,Evers13},
and, in our model, the subdiffusion emerges when the maximum force exerted by
the speckles overcomes the self-propulsion.

\begin{figure}[!t]
\centering
\includegraphics[width=.7\columnwidth]{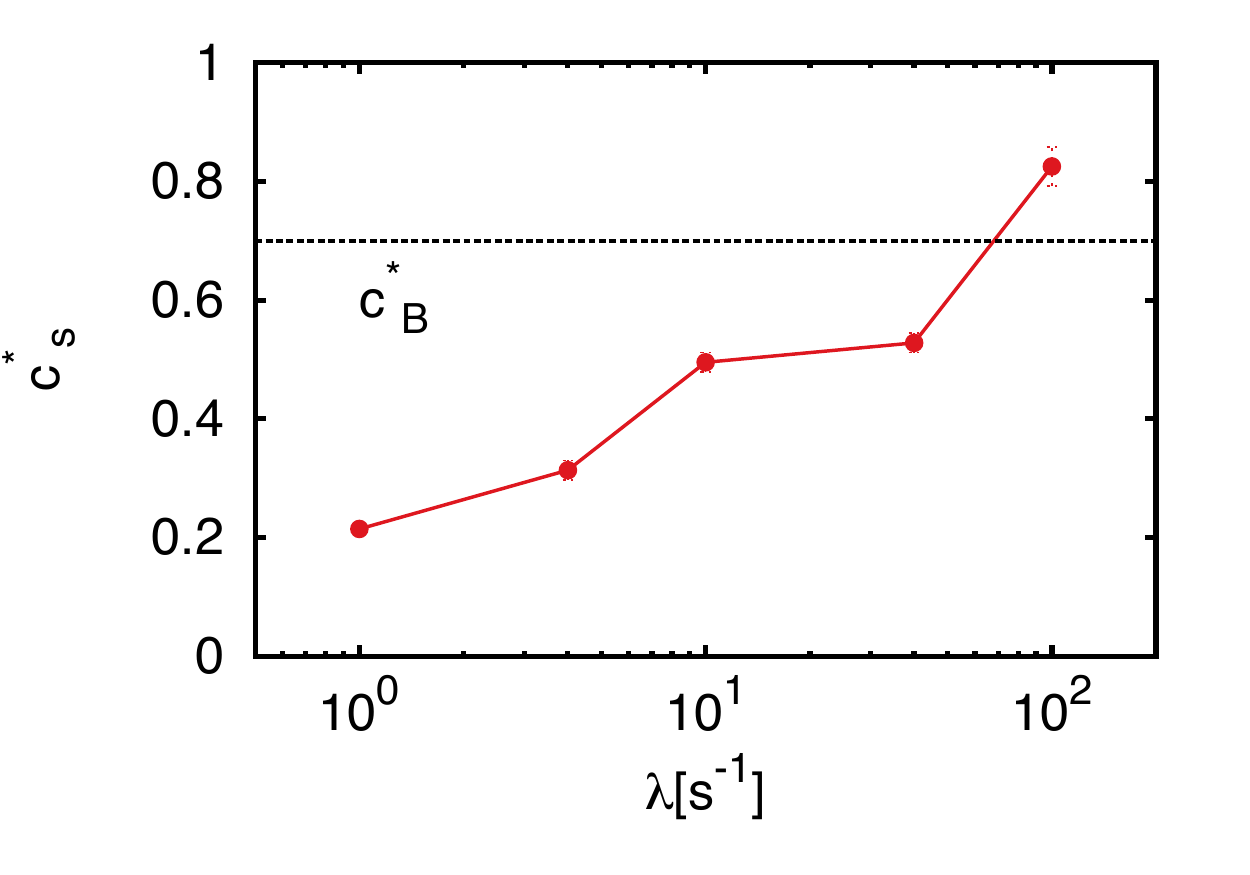}
\caption{{\bf{Threshold values as a function of the tumbling rate.}} The threshold values
are computed looking at the derivative of the entropy with respect to the control parameter $c$.
Increasing the tumbling rate (keeping the diffusion coefficient fixed) one has to increase the
intensity of the speckles to trap the particles. The dashed line is the Boltzmann limit
($\lambda\to\infty$) for non-interacting particles.}
\label{fig:plat2}       
\end{figure}
We study the Brownian limit of the run-and-tumble dynamics increasing the tumbling rate $\lambda$
and the velocity $v_0$. We perform numerical simulations
for $\lambda^{-1}=1.0, 0.25,0.1, 0.025,0.01$ s
and the self propulsion velocity is changed according to 
$v = \sqrt{2 \lambda D} \, .$
Embedding the system into the speckle pattern $U_1$ and varying $c$, we look at the entropy
to compute the threshold value $c^*_s(\lambda)$, the results being shown in Fig. (\ref{fig:plat2}).
As we can see, increasing the tumbling rate we have to increase the intensity of the speckle in
order to trap the particles. In Fig. (\ref{fig:plat2}) we also report the crossover value $c^*_B$ obtained
considering the Boltzmann limit of dilute (ideal gas) run-and-tumble particles. 
It is worth noting that for the higher $\lambda$ value one has $c^*(\lambda=100)>c^*_B$, may be due to
the excluded volume effects, not included to
estimate $c_B$.

\section{Conclusions}\label{conclusions}
We have numerically investigated steric-interacting run-and-tumble particles embedded
in random energy landscapes generated by speckle fields. 
The main result is the appearance of a crossover that separates the non-trapped to the trapped regime upon increasing speckle intensity.
The crossover value for the external field $c^*$ can be estimated from the behavior of dynamical observables, as the collective density fluctuations $F_{coll}(q,t)$ or 
static observables, as the density profiles $\rho(\rr)$, the entropy
of the density distribution $s[\rho]$ and the probability distribution of the velocity $P(v)$.
The obtained threshold $c^*$ values are of the same order of magnitude 
of $c_{max}$, i. e., the value of $c$ for which the maximum force exerted on the
system by the speckles equals the self-propulsion force of the bacterium. 
For large values of the intensity, the dynamics of the model becomes subdiffusive.
The study is performed by means of two types of patterns namely, $U_1$ ---the standard
speckle pattern--- and $U_2$, i.e., the speckle due to only
the real part of the electric field. The patterns are generated using the same
configuration of wave-vectors and phases and the fields have the same energy. From our analysis follows that pattern $U_2$ traps
more efficiently than $U_1$.

We have compared the results respect to those obtained in the Boltzmann regime.
Increasing the tumbling rate $\lambda$ and the velocity $v$ at 
fixed diffusivity $D=v^2/2 \lambda$, we have studied the Brownian
limit of the model, comparing the static properties obtained with the Boltzmann 
statistics at the effective temperature $T_{eff}$. In absence of steric interaction, 
the Boltzmann measure is concentrated in the minima of the potential. Entropy decreases
and the derivative $ds/dc$ shows a minimum at $c^*_{B}>c^*$. As a consequence,
in order to trap Brownian particles (driven by the dynamics to the Boltzmann equilibrium) we
have to increase the intensity of the speckles with respect to the case of active particles.
\ack
We acknowledge support from MIUR-FIRB project RBFR08WDBE. The research leading to these results has received funding from the European Research Council under the European Union's Seventh Framework Programme (FP7/2007-2013) / ERC grant agreement n$^\circ$ 307940. MP thanks to S. Rold\'an-Vargas for many stimulating  discussions.

\end{document}